\documentclass[preprint,5p,times,twocolumn,numbers,sort,compress]{elsarticle}
\usepackage{amssymb}
\usepackage{amsmath}
\usepackage{upgreek}
\journal{Solid State Communications}

\begin{document}

\begin{frontmatter}

\title{Hole to Electron Crossover in a (Cd,Mn)Te Quantum Well through Surface Metallization} 

\author[UW]{A. Dydniański}
\author[UW]{M. Raczyński}
\author[UW,UM]{A. Łopion}
\author[UW]{T. Kazimierczuk}
\author[Neel,JFAST,UW]{J. Kasprzak}
\author[UW,DTU]{K. E. Połczyńska}
\author[UW]{W. Pacuski}
\author[UW]{P. Kossacki}

\address[UW]{Institute of Experimental Physics, Faculty of Physics, University of Warsaw, ul. Pasteura 5, Warsaw, 02-093, Poland}
\address[JFAST]{Japanese-French Laboratory for Semiconductor Physics and Technology (J-FAST), CNRS, Universit{\'e} Grenoble Alpes, Grenoble INP, University of Tsukuba, 1-1-1 Tennodai, Tsukuba, Ibaraki 305-8573, Japan}
\address[UM]{Institute of Physics, University of Münster, Münster, 48149, Germany}
\address[Neel]{Institut N{\'e}el, CNRS, Universit{\'e} Grenoble Alpes, Grenoble INP, 38000 Grenoble, France}
\address[DTU]{DTU Electro, Technical University of Denmark, \O{}rsteds Plads 343, 2800 Kongens Lyngby, Denmark}
\begin{abstract}
In this work we look into how the contact material influences the local charge properties of a p-type CdTe-based quantum well. We study five metals deposited as 10\,nm layers on the sample surface: Au, Ag, Cr, Ni and Ti. We use magneto-spectroscopy to discriminate their charge states through monitoring the Zeeman shifts at singlet-triplet transitions. Most tested metals retain the original p-type of the QW, while gold and nickel coverage flips the local doping to n-type. This is attributed to a robust bonding of these two metals to the semiconductor, efficiently passivating its surface and thus improving electron diffusion from the metal to the quantum well.
\end{abstract}


\end{frontmatter}

\section{\label{sec:intro} Introduction}
Semiconductor quantum wells (QWs) are a canonical example of two-dimensional structures. Their excellent crystalline quality is nowadays routinely achieved thanks to the mature epitaxial growth, making them an ideal platform for optical studies of various exciton complexes. In particular, CdTe-based quantum wells were the first system in which charged excitons were reported \cite{KKhengRCox}. The incorporation of Mn$^{2+}$ ions dispersed in a II–VI semiconductor matrix enhances the effect of the magnetic field on carriers through the sp–d exchange interaction \cite{gaj_dms}. This phenomenon, characteristic of diluted magnetic semiconductors, is responsible for the emergence of giant magneto-optical effects such as giant Zeeman splitting and giant Faraday rotation \cite{gaj_faraday}. Such effects enable also efficient spin polarization of carriers.
Studies of charged exciton complexes require precise control and tuning of the carrier density in quantum wells. This can be achieved by modulation doping or by applying externally controlled electric fields. In most cases, voltage control requires fabrication of conductive contacts, in the simplest case in the form of metallic layers on the surface \cite{boukari}. (Cd,Mn)Te/(Cd,Mg)Te quantum wells with thin barrier-caps, which are the subject of this study, are known to be intrinsically p-type - the two-dimensional hole gas (2DHG) originates from surface states of (Cd,Mg)Te \cite{Maslana2003_surface}. Furthermore, the effective doping density was previously shown to be tunable by applying additional above-barrier illumination \cite{Kossacki1999}.
In this work, we examine the impact of surface metallization on the carrier gas introduced by surface states. Specifically, we investigate how the contact material influences the sign of the carriers. We compare the results obtained for five metals: Au, Ag, Ni, Ti, and Cr. Surprisingly, Au and Ni are the only metals that convert the QW doping to n-type. This demonstrates that a simple picture based solely on the metal work function is insufficient.

Our study employs a simple contactless all-optical approach to detect carriers in the quantum well. A clear fingerprint of carrier gas presence in the system is the observation of a charged exciton (X$^+$/X$^-$) line alongside the neutral exciton (X) line in photoluminescence (PL) or reflectance spectra. However, it is challenging to determine the sign of the charged exciton — i.e., the carrier gas type — by optical means alone \cite{KossackiPrzeglad}. More advanced methods are usually required. A common approach is to measure PL as a function of applied voltage in a gated sample. This, however, requires growth on a conductive substrate, which is not always feasible. In this paper we exploit an alternative technique for optical determining the sign of the charged exciton: PL measurements in a magnetic field. This method allows us to distinguish between X$^+$ and X$^-$ by monitoring the singlet–triplet transition in the studied system, and it has already been successfully employed in previous research on (Cd,Mn)Te QWs \cite{Kossacki2004,Lopion2020}.

\section{\label{sec:experiment}Samples and experiment}
The sample studied in this paper was grown by molecular beam epitaxy (MBE) and contained a single quantum well. Its structure is schematically shown in Fig.\,\ref{Figure:1}(a). On a semi-insulating GaAs (100) substrate, a 3.5\,$\upmu$m CdTe buffer and a 2\,$\upmu$m thick (Cd$_{0.77}$,Mg$_{0.23}$)Te layer were grown. The latter layer served as the lower barrier of the quantum well. Next, a 10\,nm (Cd,Mn)Te quantum well layer was deposited, followed by a 50\,nm cap-barrier of (Cd$_{0.77}$,Mg$_{0.23}$)Te.  

The manganese concentration in the QW was chosen to ensure significant Zeeman splitting while maintaining narrow excitonic lines. The exact composition was determined to be 0.03\% by fitting the modified Brillouin function \cite{gaj50}, as described in Section~4. The magnesium content in the barrier material was confirmed through reflectance measurements \cite{WaagBariera}.  

After growth, the wafer was cleaved into pieces used for characterization and processing. 
The first sample (QW1) was used to compare different metals. External hard masks were placed on the surface, through which metals were evaporated consecutively with displacement of the mask. In this way, five 10\,nm thick contact pads with $\sim$100\,$\upmu$m diameter were created on a single sample, each made of a different metal: Au, Ag, Cr, Ni, and Ti (Fig.\,\ref{Figure:1}(a)). Two other pieces, QW2 and QW3, were used to test the possible impact of the standard lithographic process. They were subjected to electron beam lithography with PMMA resist and MIBK:IPA 1:3 developer. Next, 10\,nm of metals was sputtered on the surface, with the deposited metals being Au and Cr, respectively. We found no significant modification of the results by the lithography treatment.

To perform magneto-spectroscopy, the samples were mounted on a holder integrated with an aspheric lens (NA = 0.68) on $x$-$y$-$z$ piezoelectric stages enabling micrometer spatial resolution and placed in a helium bath cryostat. The cryostat allowed measurements at temperatures down to 1.4\,K using pumped liquid helium and was equipped with superconducting coils providing magnetic fields up to 8\,T. The photoluminescence was excited using a 647\,nm CW laser. The excitation power was usually set to 1\,$\upmu$W, ensuring that no significant heating or saturation effects were present. To verify the reproducibility of the results, most measurements were performed at two temperatures: 1.8\,K and 5\,K.

\section{\label{sec:metal_impact}Impact of different metals on the optical response of the QW}
\begin{figure}[h!]
    \centering
    \includegraphics{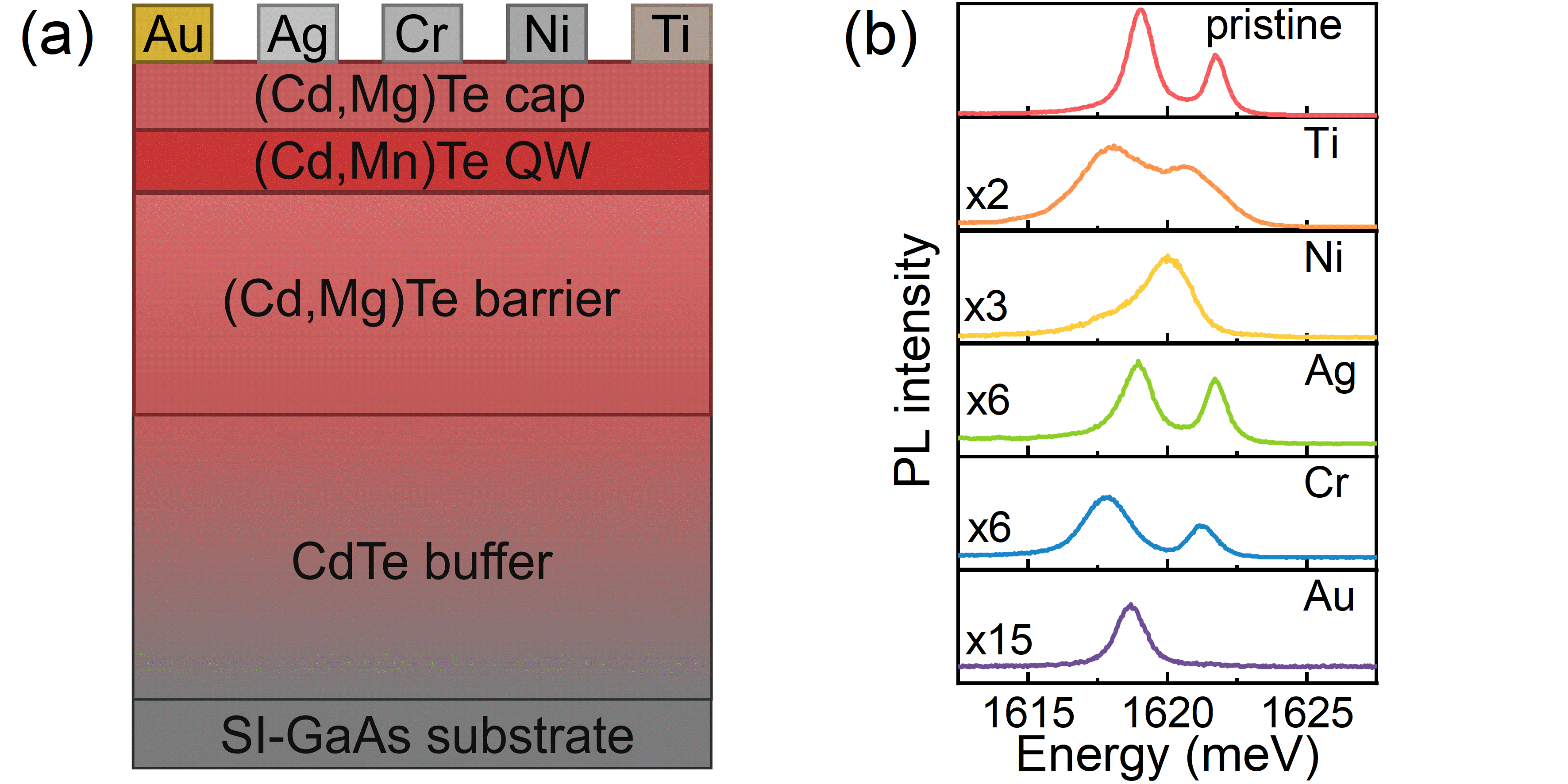}
    \caption{(a) Schematic depiction of sample QW1: 10\,nm of metals evaporated onto the quantum well surface (not in scale). (b) Photoluminescence spectra measured at metallized areas. T = 5\,K, $\lambda_{\mathrm{exc}}=647$\,nm, $P_{\mathrm{exc}}=1\,\upmu$W.}
    \label{Figure:1}
\end{figure}

All samples were subjected to preliminary $\upmu$-photoluminescence measurements. Figure~\ref{Figure:1}(b) presents photoluminescence spectra taken from metallized areas, as well as from a pristine area for comparison. As expected, the PL intensity is quenched in the regions covered by metal films due to the metals' high extinction coefficient and reflection at the air--metal boundary. Consequently, the QW underneath both receives less excitation power and emits less detectable light.  

The measured PL spectra mostly display two lines related to neutral and charged exctions. We find that spectral line shape of the emission depends on the used metal. Ti, Ni and Cr pads generate slight energy shifts, while Ti and Ni additionally cause strong broadening. The Au contact exhibits almost no neutral exciton peak, with the energy of the charged exciton remaining unshifted. Finally, virtually no modification in transition energy and line shape is observed for the Ag pad.

\section{\label{sec:magnetic-field}(Cd,Mn)Te QW in magnetic field - ascertaining the doping type by singlet-triplet transition}
The zero-field photoluminescence spectra are not sufficient to distinguish the sign of the doping, since the X--CX splitting is very similar for positively (X$^+$) and negatively (X$^-$) charged excitons \cite{KossackiPrzeglad,finkelstein1996X-,astakhov2002binding}. To infer the doping type we conducted magneto-spectroscopy.

Due to the $s,p$--$d$ exchange interaction in the (Cd,Mn)Te material, the electron and heavy-hole subbands exhibit a giant Zeeman splitting proportional to the average spin of Mn$^{2+}$ ions $\langle S^z_{\mathrm{Mn}} \rangle$. The splitting is given by  
$\Delta E_{el} = xN_0\alpha \langle S^z_{\mathrm{Mn}} \rangle$ for electrons and  
$\Delta E_{hh} = - xN_0\beta \langle S^z_{\mathrm{Mn}} \rangle$ for heavy holes.  
The exchange integrals are $N_0\alpha = 0.22$\,eV and $N_0\beta = -0.88$\,eV, respectively \cite{Gajexchange}.  
Consequently, the exciton splitting is obtained as the sum of both contributions:  
$\Delta E_X(B) = \Delta E_{el} + \Delta E_{hh}$.  

\begin{figure}[h!]
    \centering
    \includegraphics{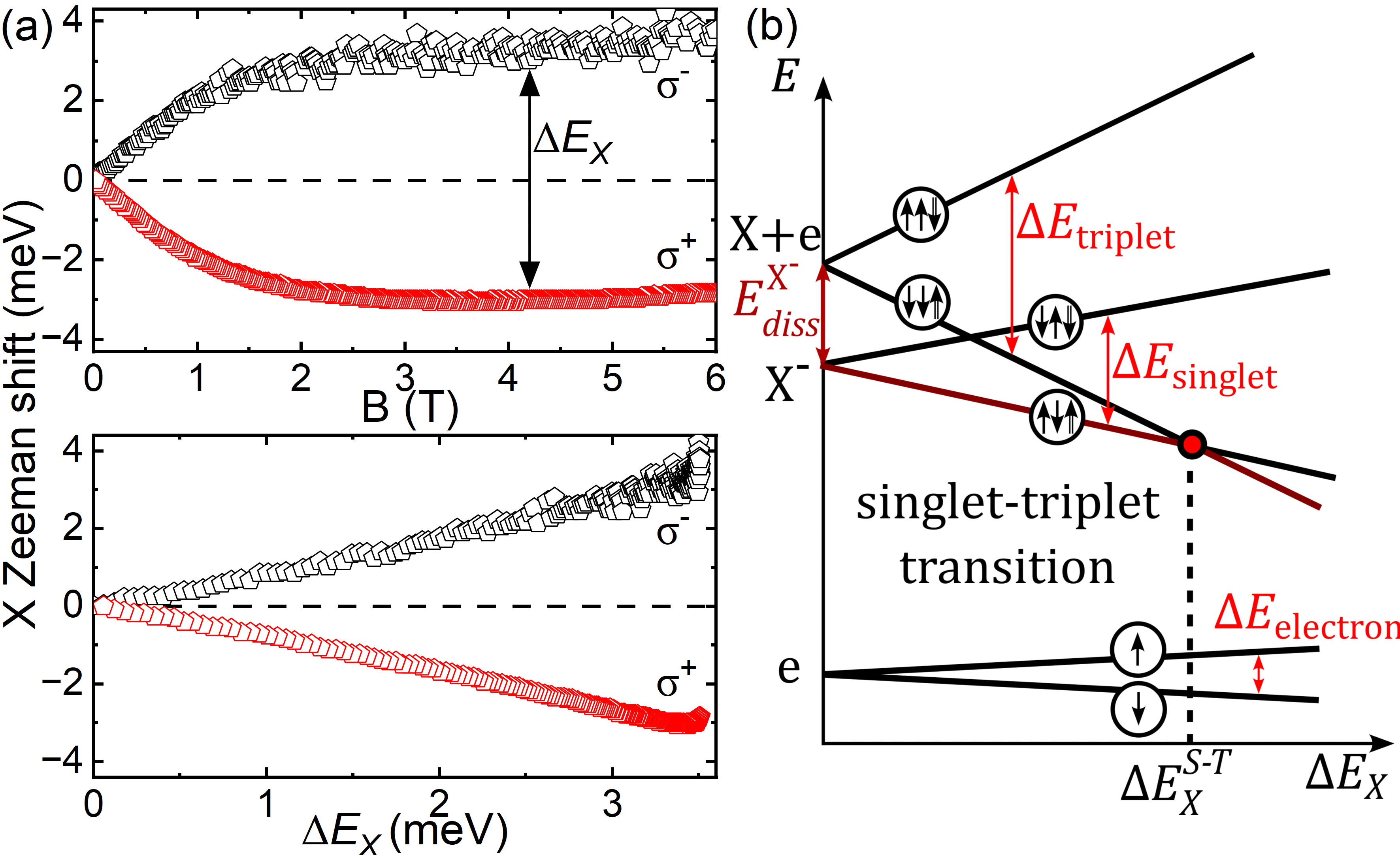}
    \caption{(a) Upper panel: giant Zeeman shift of the neutral exciton as a function of magnetic field. Exciton splitting $\Delta E_X$ is obtained as the energy difference between $\sigma^-$ and $\sigma^+$ signals. Lower panel: experimental giant Zeeman shift of the neutral exciton as a function of calculated exciton splitting $\Delta E_X$, after fitting the modified Brillouin function \cite{gaj50} to the data in the upper panel. (b) Energy levels diagram for the singlet-triplet transition in an n-type (Cd,Mn)Te QW. }
    \label{Figure:2}
\end{figure}

In magnetic field, the excitonic photoluminescence evolves into two components of opposite circular polarisations, yielding the exciton splitting as the energy difference between the $\sigma^-$ and $\sigma^+$ signals, as shown in Fig.\,\ref{Figure:2}(a). Its dependence on magnetic field and temperature follows the magnetization and can be described by the modified Brillouin function \cite{gaj_faraday}. The experimental data were fitted using the parametrization given in \cite{gaj50}, which allowed us to determine the Mn concentration and, subsequently, to calculate the Zeeman splitting $Z$ for any magnetic field and temperature. This is highly advantageous for further analysis, as $Z(\Delta E_X)$ is invariant with respect to Mn composition and temperature, allowing direct comparison of results across a range of samples and measurement conditions.  

We observe transitions associated with charged and neutral excitons (X$^+$/X$^-$ and X). To account for their energetic crossing, we followed the approach presented in \cite{Kossacki2004}. Both exciton complexes are treated as three-particle states (with a single carrier left after the radiative recombination). The singlet state of the charged exciton is bound, whereas the higher-energy unbound triplet state corresponds to the neutral exciton. In the presence of a magnetic field, all these states split.  
The splitting of the triplet state is larger than that of the singlet, and at a certain magnetic field the levels cross, making the triplet state the lower-energy one, as schematically shown in Fig.\,\ref{Figure:2}(b). Experimentally, this is evidenced by an increase in the intensity of the X line relative to the initially stronger X$^+$/X$^-$ line in photoluminescence \cite{Lopion2020,Kossacki2004}.  
This reasoning reveals two results. Firstly, the exciton splitting at the singlet–triplet transition, $\Delta E_\mathrm{X}^{S-T}$, differs for X$^+$ and X$^-$.  
Following the approach presented in previous works and using the experimental ratio $N_0\beta / N_0\alpha = 4$, we obtain:  
$   E_{\mathrm{diss}}^{\mathrm{X^+}} = \frac{4}{5} \Delta E_\mathrm{X}^{S-T}$,
 and 
$   E_{\mathrm{diss}}^{\mathrm{X^-}} = \frac{1}{5} \Delta E_\mathrm{X}^{S-T}$ respectively. Provided that the dissociation energies of X$^+$ and X$^-$ are not drastically different, the singlet–triplet transition for X$^-$ should occur at higher $B$ (or equivalently, larger $\Delta E_X$). This attribution is the key to indicate the QW doping type by magneto-spectroscopy.

\begin{figure}[h!]
    \centering
    \includegraphics{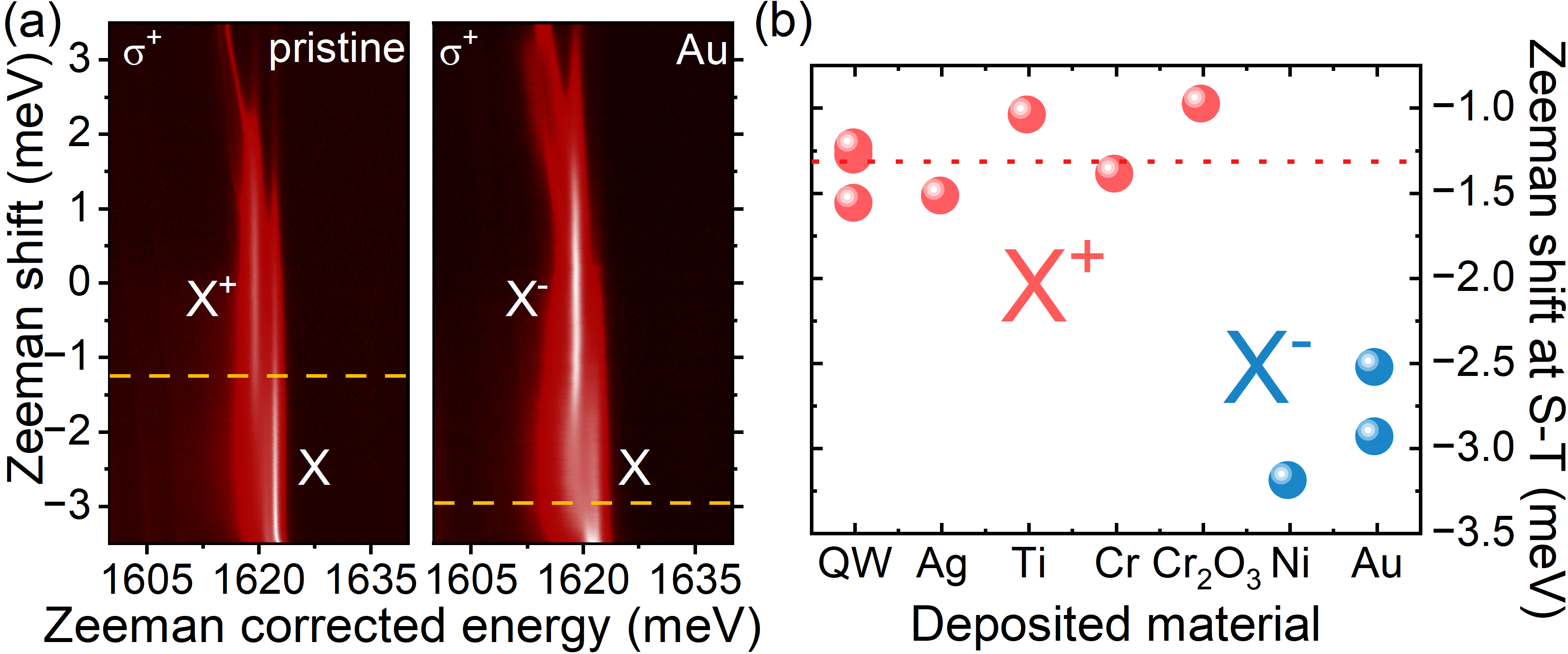}
    \caption{(a) Photoluminescence scan in magnetic field after correcting the $x$ axis for Zeeman splitting. Measurement for a pristine QW on the left, and for the metallized gold contact on the right. Orange line marks Zeeman shifts at singlet-triplet transition. (b) Complete results of Zeeman shifts at singlet-triplet transition of all materials on all samples: QW1, QW2 and QW3. Red line marks the value for a pristine QW for comparison. Cr$_2$O$_3$ was also deposited and subsequently measured. Multiple points for one material are measurements at different temperatures or on other samples.}
    \label{Figure:3}
\end{figure}

With this in mind, we turn to determining the doping type of pristine and differently metallized areas on our samples, as exemplified in Fig.\,\ref{Figure:3}(a). Measured photoluminescence was corrected for the Zeeman splitting and the singlet-triplet transition Zeeman shifts were determined by identifying points at which the intensity of the CX and X line was equal.

The pristine p-type quantum well exhibits the singlet-triplet transition at Zeeman shift of -1.3\,meV, which after converting to charged exciton dissociation energy yields $E_\mathrm{diss}^{\mathrm{X}^+}=2.1$\,meV. This value is in agreement with previous studies on p-type (Cd,Mn)Te QWs \cite{Lopion2020,Kossacki2004,dydnianski2025carrier,Kossacki1999}.

The measurement of Au-metallized area reveals that the singlet-triplet transition is at much higher Zeeman shift values of about -3.0\,meV, indicating a local change in doping from p- to n-type. Calculating the negatively charged dissociation energy yields $E_\mathrm{diss}^{\mathrm{X}^-}=1.2$\,meV. This value is smaller than the ones reported in the literature for the negatively charged exciton \cite{wojtowicz1998modulation,KossackiPrzeglad}. We speculate this discrepancy to be due to modification of electric field inside the QW by the metal on the surface - in most reported cases of X$^-$ in (Cd,Mn)Te QW the n-type doping is achieved through homogenous doping of the barrier material \cite{golnik2004ssc,konig_ntype}.

Fig.\,\ref{Figure:3}(b) assembles the results obtained on all available materials. Most metal contacts retain the original p-type of the QW. Gold and nickel are the only ones found to exhibit much larger Zeeman shifts for the singlet-triplet transition, indicating the presence of free electrons in the QW.

Our results show that our system cannot be explained with a simple model of a Schottky junction forming between the (Cd,Mg)Te and metal. As the work function of all studied metals is smaller than the work function of CdTe \cite{crchandbook,solarLi}, n-type doping was expected in all cases. Our results are corroborated by DFT-GW simulations showing that gold and nickel uniquely exhibit strong interfacial hybridization with CdTe \cite{solarLi}. This hybridization is expected to neutralize the acceptor levels on the (Cd,Mg)Te surface, enabling the Fermi level in the QW to be shifted upward. In contrast, other metals adhere weakly to (Cd,Mg)Te, leaving the surface states unpassivated and thus preserving the p-type doping.

\section{Summary}
In this study we investigated the influence of surface metallization on the charge properties of a 10\,nm p-type (Cd,Mn)Te/(Cd,Mg)Te QW. 10\,nm thin metal contacts of Au, Ag, Ni, Ti and Cr were evaporated or sputtered on the surface of the quantum well. 
The PL spectroscopy at cryogenic temperatures revealed significant spectral line shape modification on most studied metals, except for Ag, where the only change from the pristine QW is the intensity of the PL. 
Magneto-spectroscopy was employed to investigate how the metal contacts impact the charge state of the 
QW. The singlet-triplet transition present in the system was used as an indicator of majority carriers.
We find that the p-type doping of the QW persists at Ag, Ti and Cr contacts, which is explained in the literature by weak adhesion of those metals to CdTe. By properly bonding to the sample surface, Au and Ni contacts passivate it, overcoming the initial p-type doping and introducing the free electrons into the quantum well. 
Our study establishes a guidance for the choice of metal contacts in future studies of (Cd,Mg)Te quantum wells in fundamental research and in opto-electronic applications.

\section{Acknowledgements}
This work was supported by the Polish National Science
Centre under decisions DEC-2020/38/E/ST3/00364, DEC-2021/41/B/ST3/04183 and DEC-2023/51/B/ST3/01710.

\bibliographystyle{elsarticle-num-names} 
\bibliography{bibliography}
\end{document}